# Two-photon resonance single ionization of the K-shell of an atomic ion

Alexey N. Hopersky, Alexey M. Nadolinsky, Sergey A. Novikov *and* Rustam V. Koneev

*Rostov State Transport University, 344038, Rostov-on-Don, Russia*
*E-mail*: qedhop@mail.ru, amrnd@mail.ru, sanovikov@gmail.com, koneev@gmail.com

---

**Abstract.** The analytical structure and absolute values of the generalized cross-section of the two-photon resonance single ionization of the K-shell of a heavy neon-like ion of an iron atom ($Fe^{16+}$) were theoretically predicted. A pronounced resonant subthreshold structure of the generalized cross-section and the effect of destructive quantum interference of the probability amplitudes of radiation transitions into virtual excited states of *p*-symmetry have been established. The presence of a valence $2p^6$-shell in the ion backbone causes the production of an additional giant resonance of the generalized cross-section as the effect of the "reverse" hard X-ray emission of $K_\alpha$- type $1s^2(2s^2)2p^5 + \hbar\omega \rightarrow 1s(2s^2)2p^6$. A scheme of the proposed experiment with linearly polarized X-ray photons is presented to verify the theoretical results obtained.

## 1. Introduction

The creation of the X-ray Free-Electron Laser (XFEL) as a source of *hard* X-ray radiation opened up the possibility of high-precision experimental and theoretical studies of one of the fundamental processes of the microcosm – two-photon (nonlinear) single ionization of *deep* shells of atoms, atomic ions, molecules and solids (see reviews [1,2]). Studies of this process have revealed, in particular, the important role of (a) relativistic and shielding effects in the *non-resonant* generalized cross-section of the two-photon ionization of the K-shell of an atom (see [3] and references therein) and (b) nondipole (quadrupole) effects in the angular distribution of photoelectrons generated by the suprathreshold two-photon ionization of the K-shell of an atom (see [4] and references therein). In these works, as far as we know, there are no studies of the *resonance* to the threshold structure of the generalized cross-section of two-photon ionization. In the works of the authors [5–7], the first theoretical studies of the generalized cross-section of a two-photon *resonant* single ionization $1s^2$-shell of atoms (Ne, Ar), beryllium-like ($Ne^{6+}$) and helium-like ($Ne^{8+}$) atomic ions were carried out in the second order of the nonrelativistic quantum perturbation theory, taking into account only the main (leading in the infinite *complete set*) part subthreshold resonant structure of the cross-section and the effects of radial relaxation of transition states in the 1*s*-vacancy field. In this article, the theory of work [5–7] is generalized to heavy neon-like atomic ions and modified to take into account (a) the completeness of the set of virtual (intermediate) states of photoexcitation, (b) the non-trivial angular structure of the probability amplitudes of transition to final states of *d*-symmetry, and (c) destructive quantum interference of partial amplitudes of the probability of transitions. The object of the study is a neon-like ion of an iron atom ($Fe^{16+}$, ion nucleus charge $Z = 26$, configuration and ground state term $[0] = 1s^2 2s^2 2p^6 \; [^1S_0]$). The choice is due to the spherical symmetry of the ground state of the $Fe^{16+}$ ion, its availability in the gas phase [8] during experiments on the absorption of two linearly polarized XFEL photons of energy $\hbar\omega$ ($\hbar$ - Planck's constant, $\omega$ - circular frequency of the photon) by the ion trapped in the "trap" [9] and the demand for its spectral characteristics in astrophysics [10].

## 2. Theory

The probability amplitudes and the generalized cross-section of two-photon ionization are obtained in the leading second (by the number of interaction vertices) order of the nonrelativistic quantum perturbation theory. In the structure of the radiation transition operator,

$$\hat{R} = -\frac{e}{m_e c}\sum_{n=1}^{\infty}(\hat{p}_n \cdot \hat{A}_n), \tag{1}$$

for the electromagnetic field operator (in the secondary quantization representation), the dipole approximation is assumed:

$$\hat{A}_n \to \sum_{\vec{k}} \sum_{\rho=1,2} \vec{e}_{\vec{k}\rho} (\hat{a}^+_{\vec{k}\rho} + \hat{a}^-_{\vec{k}\rho}), \qquad (2)$$

$$(\vec{k} \cdot \vec{r}_n) \ll 1 \Rightarrow \exp\{\pm i(\vec{k} \cdot \vec{r}_n)\} \cong 1. \qquad (3)$$

For the applicability criterion of the dipole approximation in the form $\theta_{nl} = \lambda_\omega / r_{nl} \gg 1$ ($\lambda_\omega$ – the wavelength of the absorbed photon, $r_{nl}$ – medium radius of $nl$–ion shells) in the case of ion Fe$^{16+}$ [$r_{1s}(r_{2p}) = 0.031\,(0.123)$ Å; calculation of this work] at the energies of the absorbed photon $\hbar\omega = 8\,(6.4)$ keV have $\theta_{1s}(\theta_{2p}) \cong 50\,(16)$. In ratios (1) – (3) $e(m_e)$ – charge (mass) of an electron, $c$ – the speed of light in a vacuum, $N$ – the number of electrons in the ion, $\hat{p}_n(\vec{r}_n)$ – momentum operator (radius-vector) of $n$–ion electron, $\vec{e}_{\vec{k}\rho}(\vec{k})$ – the polarization vector (wave vector) of the photon and $\hat{a}^+_{\vec{k}\rho}$ ($\hat{a}^-_{\vec{k}\rho}$) – the photon production (annihilation) operator.

Consider the following channels of two-photon single ionization of the K-shell of a neon-like atomic ion:

$$2\omega + [0] \to 2p^5 xl + \omega \to 1s\varepsilon l, \qquad (4)$$

$$2\omega + [0] \to 1s(n,x)p + \omega \to 1s\varepsilon l. \qquad (5)$$

In (4), (5) and further the atomic system of units is accepted ($e = \hbar = m_e = 1$), $x(\varepsilon)$ – the energy of a continuous-spectrum electron, $n$ – principal quantum number of the excited state of the discrete spectrum, $l = s, d$ and filled ion configuration shells are not specified. Not included in (4) intermediate $2p^5 nl$ – discrete spectrum states at $\omega \gg I_{2pnl}$ ($I_{2pnl}$ – energy $2p \to nl$ photoexcitation) are suppressed by the energy denominator $(\omega - I_{2pnl} + i\gamma_{2p})^{-1}$ [$\gamma_{2p} = \Gamma_{2p}/2$, $\Gamma_{2p}$ – natural width of decay $2p$–vacancy] and the small value of the integral of the overlap $\langle nl | \varepsilon l \rangle \ll 1$ in the amplitude of two-photon ionization. Strong spatial and energetic distance of the subvalent ($2s^2$) and valence ($2p^6$) shells from deep $1s^2$–shells of the Fe$^{16+}$ ion [$r_{1s} = 0.031$ Å $\ll r_{2s}(r_{2p}) = 0.140\,(0.123)$ Å; $I_{1s} = 7699.23$ eV (calculation of this work) $\gg I_{2s}(I_{2p}) = 1397.77\,(1270.60)$ eV [11]; $I_{nl}$ – ionization threshold energy $nl$–ion shells] allows you to neglect the birth of finite $2s\varepsilon(s,d)$– and $2p^5\varepsilon(p,f)$–states of two-photon ionization. Finally, the probability amplitude of two-photon ionization along the channel $2\omega + [0] \to 1s\varepsilon l$ is determined by the contact interaction operator $\hat{C} = (1/2c^2)\sum_{n=1}^N (\hat{A}_n \cdot \hat{A}_n)$ and proportional to the matrix element $\langle 1s | j_l | \varepsilon l \rangle$, where $j_l$ – spherical Bessel function. In a dipole approximation for $\hat{A}_n$–operator $j_0 \to 1$, $j_2 \to 0$ and $\langle 1s | j_l | \varepsilon l \rangle \to 0$.

The probability amplitudes of two-photon ionization in channels (4), (5) are physically interpreted in Fig. 1 in the formalism of (nonrelativistic) Feynman diagrams. Let us establish their analytical structure.

**2.1. Amplitude by the channel (4)**

According to Fig. 1a for the desired amplitudes we have:

$$A_l = \sum_{M'} \int_0^\infty dx \frac{\langle 0 | \hat{R} | \Phi_{xl} \rangle \langle \Phi_{xl} | \hat{R} | \Psi_{\varepsilon l} \rangle}{\omega - I_{2p} - x + i\gamma_{2p}}, \qquad (6)$$

$$|0\rangle = [0] \otimes (\hat{a}^+_\omega)^2 |0_{ph}\rangle, \qquad (7)$$

$$|\Phi_{xl}\rangle = |2p^5 xl(^1P_1), M'\rangle \otimes \hat{a}^+_\omega |0_{ph}\rangle, \qquad (8)$$

$$|\Psi_{\varepsilon l}\rangle = |1s\varepsilon l(^1l_{J=l}), M\rangle \otimes |0_{ph}\rangle. \qquad (9)$$

In (6) – (9) the following are determined: complete wave functions of the initial ($|0\rangle$), intermediate ($|\Phi\rangle$) and final ($|\Psi\rangle$) states of two-photon ionization, projections of the total moments of the "ionic residue ⊕ electron" system $M' = 0, \pm 1$, $M = 0$ for $l = s$, $M = 0, \pm 1, \pm 2$ for $l = d$ and $|0_{ph}\rangle$ – wave function of the photon vacuum of quantum electrodynamics. Structure of the $^1l_{J=l}$ –term of the final states of two-photon ionization ($J = 0 \Rightarrow {}^1S_0$; $J = 2 \Rightarrow {}^1D_2$) reproduces the Landau–Yang theorem [12,13] for the total angular moment of a system of two absorbed photons $J_\omega = 0, 2$. Using the methods of algebra of photon production (annihilation) operators, the theory of irreducible tensor operators, the theory of nonorthogonal orbitals (see, e.g., [14] and references therein) and approximation for the integral of the overlapping wave functions of the continuous spectrum $\langle xl | \varepsilon l \rangle \cong \delta(x - \varepsilon)$ [$\delta$ – Dirac's delta function], for (6) get:

$$A_s = \xi \langle 2p_0 | \hat{r} | \bar{\varepsilon} s_+ \rangle, \qquad (10)$$

$$A_d = \sqrt{6}\, \xi \langle 2p_0 | \hat{r} | \bar{\varepsilon} d_+ \rangle \cdot Q_M, \qquad (11)$$

$$\xi = \frac{4\pi}{3V\omega} \frac{\omega_{sp}(2\omega - \omega_{sp})}{(\omega - \omega_{sp} - i\gamma_{2p})} \langle 1s_0 | \hat{r} | 2p_+ \rangle, \qquad (12)$$

$$Q_M = -\frac{4\pi}{3} \sum_{M'} \sum_p (-1)^{M'} Y_{1,M'}(\vec{e}_\omega) \cdot Y_{1,p}^*(\vec{e}_\omega) \begin{pmatrix} 1 & 1 & 2 \\ -M' & p & M \end{pmatrix}. \qquad (13)$$

In here $\bar{\varepsilon} = 2\omega - I_{1s}$, $\omega_{sp} = I_{1s} - I_{2p}$, $V(cm^3) = c$ – the volume of quantization of the electromagnetic field (*numerically* equal to the speed of light in a vacuum) [15], $Y_{\alpha,\beta}(\vec{e}_\omega)$ – spherical function, $\vec{e}_\omega$ – the polarization vector of the absorbed photon, $p = 0, \pm 1$, «*» – symbol of complex conjugation and defined of the 3$j$-Wigner symbol. In equations (10) – (12) indices «0» and «+» correspond to the radial parts of the wave functions of electrons obtained by solving the single-configuration equations of the self-consistent Hartree-Fock field for [0]– and $1s_+\varepsilon l_+$ –configurations of ion states. Multiplier spawn $Q_M$ in (11) reflects the fact of a non-trivial angular structure of the amplitude of the probability of transition to final states $d$ –symmetry (in the top line 3$j$-Wigner symbol $J = 2$).

## 2.2. Amplitude by the channel (5)

According to Fig. 1b for the desired amplitudes we have the quantum interference of partial amplitudes:

$$B_l = B_l^{(1)} + B_l^{(2)}, \qquad (14)$$

$$B_l^{(1)} = \sum_{M'} \sum_{n=3}^{\infty} \frac{\langle 0 | \hat{R} | K_n \rangle \langle K_n | \hat{R} | \Psi_{\varepsilon l} \rangle}{\omega - I_{1snp} + i\gamma_{1s}}, \qquad (15)$$

$$B_l^{(2)} = \sum_{M'} \int_0^\infty dx \frac{\langle 0 | \hat{R} | K_x \rangle \langle K_x | \hat{R} | \Psi_{\varepsilon l} \rangle}{\omega - I_{1s} - x + i\gamma_{1s}}, \qquad (16)$$

$$|K_{n,x}\rangle = |1s(n,x)p\,({}^1P_1), M'\rangle \otimes \hat{a}_\omega^+ |0_{ph}\rangle, \qquad (17)$$

where $I_{1snp}$ – energy $1s \to np$ photoexcitation and $\gamma_{1s} = \Gamma_{1s}/2$, $\Gamma_{1s}$ – natural width of decay $1s$ – vacancy. Following the methods of construction $A_l$ –amplitudes of the previous Section 2.1 and taking the approximation of plane waves [$|x(r)\rangle \sim \sin(r\sqrt{2x})$] for the single-electron amplitude of the probability of the radiation transition between continuous-spectrum states in (16),

$$(x - \varepsilon)\langle xp_+ | \hat{r} | \varepsilon d_+ \rangle \cong i\sqrt{2x} \cdot \delta(x - \varepsilon), \qquad (18)$$

for (14) get:

$$B_s = \eta(\mu + \sum_{n=3}^{\infty} \beta_n \langle np_+ | \hat{r} | \bar{\varepsilon} s_+ \rangle), \tag{19}$$

$$B_d = \sqrt{6} \cdot \eta(\mu + \sum_{n=3}^{\infty} \beta_n \langle np_+ | \hat{r} | \bar{\varepsilon} d_+ \rangle) \cdot Q_M, \tag{20}$$

$$\mu = i \cdot 2\sqrt{2\bar{\varepsilon}} \langle 1s_0 \| \hat{r} \| \bar{\varepsilon} p_+ \rangle, \tag{21}$$

$$\beta_n = \frac{I_{1snp}(2\omega - I_{1snp})}{(\omega - I_{1snp} + i\gamma_{1s})} \cdot \langle 1s_0 \| \hat{r} \| np_+ \rangle, \tag{22}$$

where $\eta = \frac{4\pi}{3} \frac{1}{V\omega}$ and the single-electron amplitude of the probability of $1s \to np$ photoexcitation was determined (taking into account the effect of radial relaxation of excited states in the Hartree-Fock field of the $1s$-vacancy by the methods of the theory of nonorthogonal orbitals):

$$\langle 1s_0 \| \hat{r} \| np_+ \rangle = N_{1s}(\langle 1s_0 | \hat{r} | np_+ \rangle - F_n), \tag{23}$$

$$N_{1s} = \langle 1s_0 | 1s_+ \rangle \langle 2s_0 | 2s_+ \rangle^2 \langle 2p_0 | 2p_+ \rangle^6, \tag{24}$$

$$F_n = \frac{\langle 1s_0 | \hat{r} | 2p_+ \rangle \langle 2p_0 | np_+ \rangle}{\langle 2p_0 | 2p_+ \rangle}. \tag{25}$$

## 2.3. Generalized ionization cross-section

Following the definition of the concept of a *generalized* cross-section of a two-photon single ionization of an atom (atomic ion) [16],

$$d\sigma_g^{(l)} = (V/2c) \cdot d\sigma_l, \tag{26}$$

considering quantum interference $A_l$– and $B_l$–amplitudes in Fermi's "golden rule" [17],

$$d\sigma_l = (\pi V/c) |A_l + B_l|^2 \delta(\varepsilon - \bar{\varepsilon}) d\varepsilon, \tag{27}$$

and integrating into (27) by the energy of the photoelectron, for the desired complete generalized cross-section, we obtain (the probability of photon disappearance *without* photoelectron registration):

$$\sigma_g (cm^4 \cdot s) = \chi \frac{1}{\omega^2} \sum_{l=s,d} \sum_{i=1,2} a_l L_{il}^2, \tag{28}$$

$$L_{1l} = (\omega - \omega_{sp}) L_l + \sum_{n=3}^{\infty} (\omega - I_{1snp}) C_{ln}, \tag{29}$$

$$L_{2l} = \gamma_{2p} L_l - \gamma_{1s} \sum_{n=3}^{\infty} C_{ln} + D, \tag{30}$$

$$L_l = \frac{\omega_{sp}(2\omega - \omega_{sp})}{(\omega - \omega_{sp})^2 + \gamma_{2p}^2} \langle 1s_0 | \hat{r} | 2p_+ \rangle \langle 2p_0 | \hat{r} | \bar{\varepsilon} l_+ \rangle, \tag{31}$$

$$C_{ln} = \frac{I_{1snp}(2\omega - I_{1snp})}{(\omega - I_{1snp})^2 + \gamma_{1s}^2} \cdot \langle 1s_0 \| \hat{r} \| np_+ \rangle \langle np_+ | \hat{r} | \bar{\varepsilon} l_+ \rangle, \tag{32}$$

$$D = 2\sqrt{2\bar{\varepsilon}} \cdot \langle 1s_0 \| \hat{r} \| \bar{\varepsilon} p_+ \rangle, \tag{33}$$

where $\chi = 0.278 \cdot 10^{-52}$ (cm$^4 \cdot$s), $a_s = 1$, $a_d = \frac{3}{2} a_d^{(0)} \left(1 - \frac{1}{4\pi}\right)$ and $a_d^{(0)} = 4/5$ [5–7] (only projection $M = 0$ total moment $J = 2$ is taken into account). To calculate the coefficient

$$a_d = 6 \sum_{M=-2}^{2} |Q_M|^2 , \qquad (34)$$

in the sum of the squares of the probability amplitudes of transitions along $M$–projections of finite states $d$–symmetry, the analytical result of the work [18] for the sum of the products of 3$j$-Wigner symbols is taken into account (see **Appendix**) and implemented the scheme of the *proposed* XFEL-experiment for linearly polarized absorbed photons: $\vec{k} \in OZ$, $\vec{e}_\omega \in OX$ ($OX, OZ$ – rectangular coordinate system axes),

$$Y_{1,0}(\vec{e}_\omega) = 0, \quad Y_{1,\pm 1}(\vec{e}_\omega) = \mp 3/(4\pi\sqrt{2}). \qquad (35)$$

According to (34), additional accounting (see $a_d^{(0)} \to a_d$) projections $M = \pm 1, \pm 2$ full moment $J = 2$ noticeably (~ 30 %) increases the contribution of the generalized cross-section (28) component to $l = d$.

## 3. Results and discussion

The calculation results are shown in Fig. 2 and Tables 1, 2. For the parameters of the generalized cross-section (28), the following values are taken: $\Gamma_{1s}$ = 1.046 eV [21], $\Gamma_{2p}$ = 0.023 eV [extrapolation of work data [22,23] for radiative decay widths $2p^5 n(s,d) \to 2p^6$, $n \in [3;\infty)$], $I_{1s}$ = 7699.23 eV (relativistic calculation of this work), $I_{2p}$ = 1270.60 eV [11]) and $\omega \in (6; 8)$ keV ([24] LCLS XFEL, USA; [25] PAL–XFEL, Republic of Korea; [26] European XFEL, Germany).

The results in Fig. 2 and Table 1 demonstrate a pronounced *subthreshold resonance* structure of the generalized cross-section of the two-photon ionization of the ion $Fe^{16+}$ at $\omega \in (6.25; 7.70)$ keV. Structure at $\omega \in (7.00; 7.70)$ keV is driven by virtual states $1s \to np$ photoexcitation (Fig. 1b; the values of the main quantum number $n \in [3; 150]$ are taken into account). Giant resonance of a generalized cross-section at $\omega = 6.4329$ keV due $1s^2 2p^5 + \omega \to 1s 2p^6$ radiation absorption of the second photon incident on the ion (Fig. 1a). Its magnitude $\sigma_g \cong 2\cdot 10^{-48}$ (cm$^4 \cdot$s) almost an order of magnitude less than the value of the leading resonance $1s \to 3p$ photoexcitation, but significantly exceeds the magnitude of resonances $1s \to mp$ photoexcitation for $m \geq 4$ (Table 1). The results in Fig. 2 also demonstrate the effect of *destructive* (damping) quantum interference of probability amplitudes of resonant states $1s^2 2p^5 \to 1s 2p^6$ and $1s \to np$ transitions. This effect is due to the alternation of the $(\omega - \omega_{sp})$– and $(\omega - I_{1snp})$– multipliers in (29) and the "immersion" of these states in the continuum [see $D$ in (30)]. At the same time, "windows of transparency" appear between the maxima of resonances of the generalized cross-section as a sharp drop in the probability of two-photon ionization of the ion $Fe^{16+}$. Formally, mathematically infinite sums of (29), (30) correspond to taking into account the *completeness of the set* of virtual states $1s \to np$ photoexcitation. Analytical calculation of these amounts is not possible. Numerical methods of summation are inevitable. This article implements the authors' method of work [27]. Values $I_{1snp}$ and $J_n = \langle 1s_0 \| \hat{r} \| np_+ \rangle$ for $n \in [3;10]$ obtained in the single-configuration Hartree-Fock approximation. For $n \in [11;\infty)$ energy $1s \to np$ excitations are obtained by approximation of the form:

$$I_{1snp} = I_{1s} - \frac{1}{n^2}\left(a - \frac{b}{n}\right), \quad \lim_{n \to \infty} I_{1snp} = I_{1s}, \qquad (36)$$

where magnitudes $a$ and $b$ defined by values $I_{1smp}$ for $m = 9,10$. For $n \in [11;\infty)$ probability amplitudes $1s \to np$ excitations are obtained by approximation of the form:

$$J_n = \frac{1}{n^2}\left(c + \frac{d}{n} + \frac{f}{n^2}\right), \quad \lim_{n \to \infty} J_n = 0, \qquad (37)$$

where magnitudes $c$, $d$ and $f$ defined by values $J_m$ for $m = 8,9,10$. For the integral $\langle np_+ | \hat{r} | \bar{\varepsilon} l_+ \rangle$ formula (37) is implemented in (32) taking into account the fact that the $c$, $d$ and $f$ become functions of the energy of the absorbed photon (see $\bar{\varepsilon} = 2\omega - I_{1s}$). As might be expected, due to inequalities for the single-electron probability amplitudes of the transition in (31) and (32)

$$\left|\left\langle 2p_0(np_+)\left|\hat{r}\right|\bar{\varepsilon}d_+\right\rangle\right| > \left|\left\langle 2p_0(np_+)\left|\hat{r}\right|\bar{\varepsilon}s_+\right\rangle\right|, \tag{38}$$

results in Table 2 show the *leading* role of the *d*–symmetry of the final ionization state in determining the value of the total generalized cross-section at $\omega \in (6; 8)$ keV. A similar statement (on the example of ionization of the K-shell of neutral atoms Ne and Ge) about the leading role of the $s \to p \to d$ channel of two-photon ionization over the $s \to p \to s$ channel is given in [3]. Thus, when the K-shell of a neon-like atomic ion is ionized once by two photons, the total moment of the photon system is most likely to be realized $J_\omega = 2$. This result is similar to that for a single-photon ($J_\omega = 1$) single ionization of the $nl$–shell of an atom (atomic ion), where the transition $l \to l+1$ called "main" [28].

## 4. Conclusion

A nonrelativistic version of the quantum theory of the process of two-photon resonant single ionization of the K-shell of a heavy neon-like atomic ion has been constructed. The effects of (a) the occurrence of giant resonances in the subthreshold region of the generalized ionization cross-section, (b) destructive quantum interference of partial amplitudes of the probability of transitions, and (c) the leading role of the *d*-symmetry of the final ionization state in determining the total generalized cross-section in the region of photon energies of the hard X-ray range $\omega \in (6; 8)$ keV. Going beyond the dipole approximation for the $\hat{R}$–operator of the radiation transition and taking into account correlation and relativistic effects is the subject of future development of the theory. The generalization of the presented theory to atoms and atomic ions of other types and the establishment of the role of the charge of their nucleus is the subject of future research. Finally, the results of successful experiments on the observation of two-photon ionization of atoms, molecules, and solids [1,2] suggest that the absolute values of the generalized cross-section in Fig. 2 are quite measurable in the modern XFEL-experiment. It is interesting to note that the "*analogue*" of the two-photon single ionization of an atom (atomic ion) in quantum electrodynamics, the *linear* Breit-Wheeler effect (the production of an electron-positron pair by two *real* $\gamma$ – quanta: $\gamma\gamma \to e^+e^-$) [29], has not yet been experimentally *unambiguously* identified [30]. However, this effect is recorded experimentally through the production of two *virtual* $\gamma$ – quanta in the peripheral collision of heavy gold ions (Au; Z=79) [31] and heavy lead ions (Pb; Z=82) [32].

---

## Appendix

Expression (34) contains the sum of $3j$-Wigner's symbol products of the form:

$$W_{cd}^{ab} = \sum_{m=-2}^{2} \begin{pmatrix} 1 & 1 & 2 \\ a & b & m \end{pmatrix} \begin{pmatrix} 1 & 1 & 2 \\ c & d & m \end{pmatrix}. \tag{A1}$$

The result of analytical summation in (A1) is established in [18]. Here we will briefly reproduce this result. Let us consider the orthogonality condition of $3j$-Wigner's symbols [19]:

$$\sum_j \sum_m (2j+1) \begin{pmatrix} j_1 & j_2 & j \\ a & b & m \end{pmatrix} \begin{pmatrix} j_1 & j_2 & j \\ c & d & m \end{pmatrix} = \delta_{a,c} \cdot \delta_{b,d}, \tag{A2}$$

where $\delta_{\alpha,\beta}$ - symbol of Kronecker-Weierstrass and the equations $a+b+m=0$, $c+d+m=0$ are fulfilled. Let us take into account the particular values of $3j$- Wigner's symbols [20]:

$$\begin{pmatrix} 1 & 1 & 0 \\ a & b & 0 \end{pmatrix} = \frac{1}{\sqrt{3}}(-1)^{1+b} \delta_{a,-b}, \tag{A3}$$

$$\begin{pmatrix} 1 & 1 & 1 \\ a & b & 0 \end{pmatrix} = \frac{1}{\sqrt{6}}(-1)^b \cdot a \cdot \delta_{a,-b}, \tag{A4}$$

$$\begin{pmatrix} 1 & 1 & 1 \\ a & b & \pm 1 \end{pmatrix} = \pm \frac{1}{2\sqrt{3}}(-1)^b \cdot \sqrt{(1 \mp b)(2 \pm b)} \cdot \delta_{a,\mp 1-b}. \tag{A5}$$

Then, with $j_1 = j_2 = 1$ and $j = 0, 1, 2$ from (A2) get:

$$W_{cd}^{ab} = \frac{1}{5}\delta_{a,c}\delta_{b,d} - \frac{1}{20}(-1)^{b+d}(E+W), \tag{A6}$$

$$E = \left(2bd + \frac{4}{3}\right)\cdot \delta_{a,-b}\delta_{c,-d}, \tag{A7}$$

$$W = W_+ + W_-, \tag{A8}$$

$$W_\pm = \sqrt{(1\pm b)(2\mp b)(1\pm d)(2\mp d)}\cdot \delta_{a,\pm 1-b}\delta_{c,\pm 1-d}. \tag{A9}$$

In our case [see (13) for $Q_M$] $a = -M'$, $b = p$, $c = -M''$, $d = p' = 0, \pm 1$ and $m = M$.

**Table 1.** Spectral characteristics of $1s \to np$ the leading resonances of the generalized cross-section of the two-photon single ionization of the K-shell of the $Fe^{16+}$ ion in the energy region of the absorbed photon $\hbar\omega \in (7.00; 7.70)$ keV. $[n] \equiv 10^n$.

| $np$ | $I_{1snp}$ (keV) | $\sigma_g$ ($10^{-53}$ cm$^4\cdot$s) |
|---|---|---|
| $3p$ | 7.194 | 6.55·[6] |
| $4p$ | 7.427 | 5.52·[3] |
| $5p$ | 7.529 | 3.48·[2] |
| $6p$ | 7.583 | 1.19·[2] |

**Table 2.** Relative contribution $s-$ and $d-$ symmetries of the final ionization state $\Lambda = \sigma_g^{(d)}/\sigma_g^{(s)}$ [see (28) for the terms $l = d$ and $l = s$] into the complete generalized cross-section of the two-photon resonant single ionization of the K-shell of the $Fe^{16+}$ ion.

| $\hbar\omega$ (keV) | 6.0 | 7.0 | 8.0 |
|---|---|---|---|
| $\Lambda$ | 2.504 | 2.573 | 2.660 |

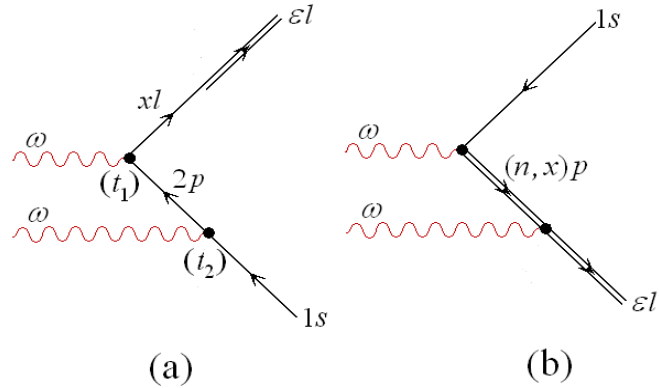

**Figure 1.** Probability amplitudes of two-photon resonant single ionization of the K-shell of a neon-like atomic ion in the representation of Feynman diagrams: (a) by the channel (4); (b) by the channel (5). Time direction – left to right ($t_1 < t_2$). An arrow to the right is an electron, an arrow to the left is a vacancy. Double line – the state is obtained in the Hartree-Fock field of $1s-$ vacancies. The connection of the single and double lines corresponds to the overlap integral $\langle xl | \varepsilon l \rangle$, $l = s, d$. The black circle corresponds to the top of the radiation transition.

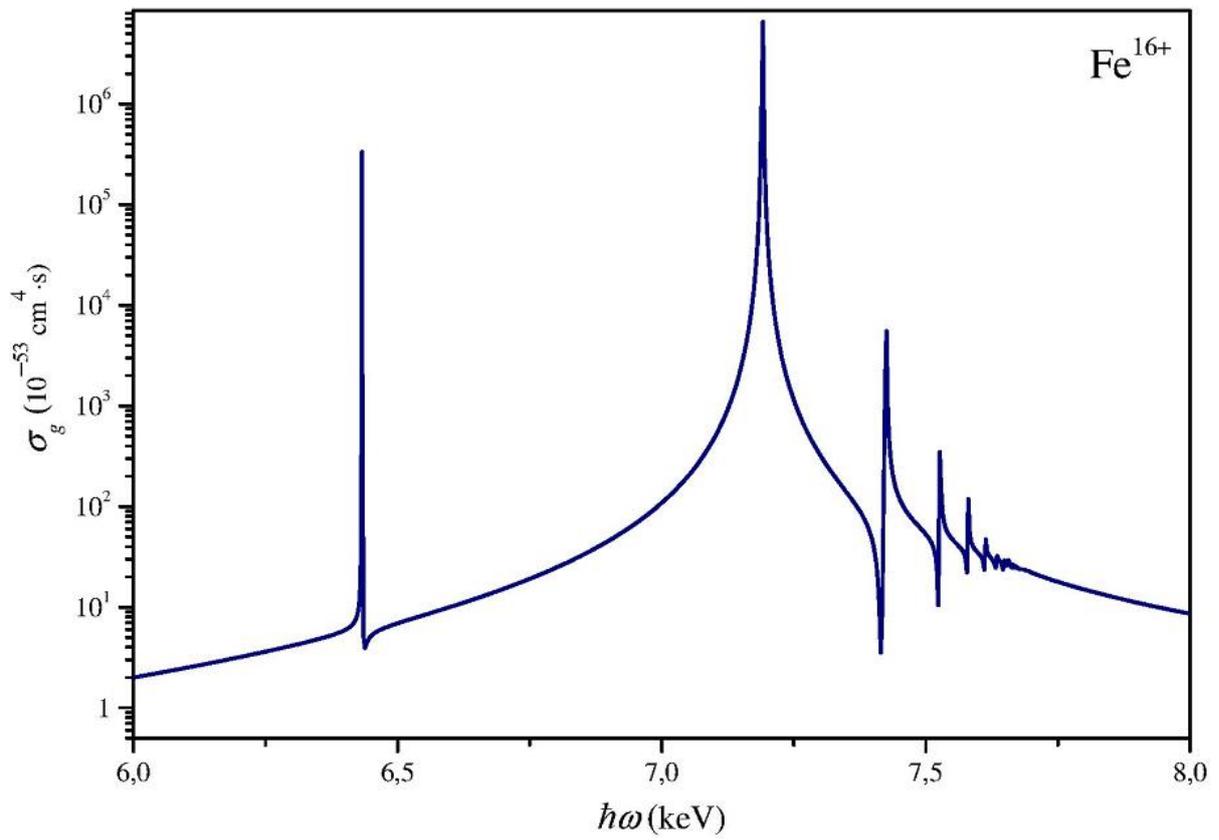

**Figure 2.** The total generalized cross-section of the two-photon resonant single ionization of the K-shell of the Fe$^{16+}$ ion. $\hbar\omega$ – the energy of the absorbed photon.